# Agentic AI Containment Architecture for Security Hardening

Mohamed ElBendary

FasTrak SoftWorks, Mequon, WI, USA
me@prosterk.com

**Abstract.** Multi-agent AI systems are increasingly deployed in contexts where autonomous coordination, tool use, and continuous learning introduce novel security and governance risks. The containment approach to multi-agent system security remains underdeveloped, primarily resorting to add-on layers post-design and sometimes post-implementation and deployment. This paper proposes an Agent Containment Architecture that treats security as an architectural property enforced through a set of explicit constraints that bound the design space of multi-agent systems. The architecture proposed in this paper introduces a novel formal organizational mapping from standard systems analysis artifacts to machine-verifiable contracts under a proposed constraint system. The architecture introduces six interacting constraints: separation of responsibility assignments, pre-deployment coherence checking, value stream binding, temporal isolation, strict knowledge verification before accumulation, and deterministic verification of structural and process integrity. Together, these constraints enforce a Propose–Verify–Act–Verify execution model in which all operations are contractually defined, independently verified, and traceable to specific execution contexts. The paper presents propositions and correctness reasoning arguments linking these constraints to defenses against key threat classes, including prompt injection, orchestrator manipulation, cross-session state poisoning, and emergent agent collusion. A resume screening case study demonstrates how the architecture produces auditable, policy-compliant outcomes under adversarial conditions. The work explicitly separates structural integrity from semantic safety, bounding residual risks while making residual semantic risk explicit and measurable.

## 1 Introduction

Multi-agent AI systems are increasingly autonomous, challenging meaningful human oversight in two critical ways. First, potential for skill atrophy over time as human verifiers are marginalized to the periphery instead of deep operational knowledge. Second, human oversight does not scale to match the execution speed and opaque reasoning, especially when continuous learning and interactions drive emergent be-havior. Under such conditions, the security of a multi-agent system must be treated as an architectural concern, not a bolt-on after development or deployment.

This work proposes a logical architecture for multi-agent systems that hoists security into the architecture. We introduce a constraint-driven architecture in which security properties are not additive features but emergent outcomes of enforced structural boundaries. By mapping specific architectural constraints to classes of adversarial behavior, the work provides a more grounded foundation for reasoning about containment in multi-agent AI systems. The architectural constraints bound the solution space of possible designs in compliance with this architecture. We also propose an organizational mapping of systems analysis artifacts onto runtime-verifiable behavioral contracts for context-based governance of AI agents.

This paper explicitly separates semantic safety from structural integrity, delegating the former to hardened inference layers while enforcing the latter through deterministic, contract-bound verification mechanisms. The proposed architecture also draws on established principles from software engineering and distributed systems. Design-by-Contract [6] provides a foundation for specifying preconditions, postconditions, and invariants governing component interactions. Similarly, domain-driven design [4] and use-case modeling [2] offer methodologies for capturing system behavior in terms of actor goals and interaction flows. In distributed systems, zero-trust architectures and Byzantine fault tolerance models emphasize the importance of independent verification and the assumption of adversarial components within a system. These paradigms inform the separation of actor and verifier roles, as well as the requirement for deterministic validation of structural properties.

The semi-formal proposition structure presented in this work provides system-level correctness arguments, not formal proofs, under explicit architectural assumptions, bridging the gap between purely informal design patterns and fully formalized systems.

## 2 Related Work

Multi-agent AI systems have rapidly evolved from experimental constructs into operational architectures capable of decomposing and executing complex tasks across heterogeneous tools and environments. Agent frameworks and SDKs offer orchestrated agent stacks that are rapidly evolving into de facto patterns of task decomposition, tool invocation, and iterative refinement [1]. These systems demonstrate emergent capability through coordination, but they largely rely on centralized orchestration logic and implicit trust assumptions between agents. As a result, they exhibit well-documented vulnerabilities, including prompt injection, tool misuse, and uncontrolled state propagation. Existing mitigation strategies in these ecosystems tend to be implementation add-on layers such as prompt hardening, sandboxing, or heuristic filtering, rather than structural. This work departs from that paradigm by treating security as an architectural property, enforced through explicit constraints on agent interaction, state management, and execution traceability.

The AI safety and alignment research has introduced techniques aimed at constraining model behavior, including Reinforcement Learning from Human Feedback (RLHF), Constitutional AI, Process-based Supervision, and rule-based guardrails at inference time. These approaches improve behavioral alignment at the model level, however, they operate primarily within the semantic layer of individual model outputs. They do not provide guarantees about system-level properties such as cross-agent accountability, execution traceability, or resistance to adversarial coordination across components. Multi-agent systems research has shown that alignment by individual models and compliance by individual agents do not guarantee compliance of an agent swarm [1, 7]. Recent work on tool-use safety and structured prompting partially addresses these gaps by introducing schema-constrained interactions and validation layers. However, these mechanisms remain embedded within probabilistic systems and are therefore susceptible to adversarial manipulation or drift. Prior work in these domains does not address the unique challenges introduced by adaptive, probabilistic agents operating across shared execution contexts.

Formal methods and verification techniques have long sought to provide guarantees about system correctness through mathematical specification and proof. Model checking, type systems, and formal verification frameworks offer strong assurances but often struggle with scalability and applicability in systems involving probabilistic components such as large language models [8]. Recent efforts in AI verification focus on bounded guarantees, adversarial robustness, or risk-constrained deployment, but do not address multi-agent interaction as a first-class verification problem [5].

Finally, emerging work in AI security has cataloged a growing taxonomy of threats, including prompt injection, data exfiltration through tool use, reward hacking, and long-term memory poisoning [4, 5]. Industry guidance increasingly recommends isolation mechanisms, audit logging, and least-privilege access controls. While these recommendations align directionally with secure system design, they are typically implemented as layered defenses rather than as constraints that shape the design space itself.

The related work surveyed here reveals a critical gap, that no existing approach treats security as a property that bounds the design space itself. The architecture proposed in this paper addresses this gap directly. By introducing six interacting constraints that collectively enforce a Propose–Verify–Act–Verify execution model, the architecture makes security properties an emergent structural outcome rather than a reactive defensive layer. Critically, it also resolves a persistent conflation in prior work by drawing a distinguishing line between structural integrity, which can be verified deterministically, and semantic safety, which cannot. By explicitly separating these two concerns and assigning each to the appropriate verification mechanism, the architecture reduces the attack surface of the verification path itself while making residual semantic risk explicit, bounded, and measurable.

# 3 Agentic Containment Architecture

## 3.1 Software Engineering Origins

In a multi-agent architecture, one or more agents carry out the orchestrator role, while other agents are assigned specific responsibilities. These responsibilities map to sub-problems forming a decomposition of the original problem submitted to the system. An AI agent or a human user may carry out the decomposition.

In software engineering, *Design-by-Contract* is a software design method that uses contract-like specifications to govern inter-component interactions. When these technical contracts are violated, a range of design options include triggering predetermined responses (logging, escalation to human oversight, or termination), and generating audit evidence that may be relevant to legal proceedings. The contracts serve as control abstractions, focusing on interface semantics while hiding implementation details needed to fulfill the contracts.

In software requirements engineering, domain modeling is an analysis activity to discover the entities, their attributes, their behaviors, and cross-entity relationships and interaction semantics. Domain operations exist organically (native to the business process regardless of technology) and are modeled directly within the system's behavior. For example, reviewing a resume or a loan application operations existed long before computers were invented.

An artifact of the domain modeling process is describing the user-system interactions from an initial state to a goal state within a particular context. The fully-dressed sea-level use cases style, as described by [2], covers actors, their goals, user-system interactions on the happy path and alternate branches, preconditions, success and failure end conditions, and escalation semantics. This amounts to a behavioral modelling framework for expressing aligned system behavior in both user-facing and machine-to-machine contexts. This use case specification expresses semi-formally what constitutes acceptable (aligned) behavior regardless of whether the interaction is human-agent or agent-to-agent.

Step-by-step modelling of happy paths and alternate paths is expressed in terms of the operations that must be carried out by interacting parties for the choreography of the interaction to succeed. This forms the basis of domain operation discovery with traceability to existing processes, regardless of current implementation technologies. The domain operations powering the interaction can be contractually modeled with respect to their required entry state(s), inputs, outputs, exceptional conditions, and escalation paths, keeping them composable and maintainable as long as their interface contracts remain stable.

## 3.2 Logical Architecture and Constraints

We propose that the agentic AI security stack decomposes into four layers, each addressing a distinct class of containment concerns. The *infrastructure layer* provides agent identity and trust management with non-repudiation and tamper-resistant logging. The *role-structural layer* bounds the decision and action space of each agent.

The *resource-access layer* enforces least privilege on what authenticated agents may reach, through sandboxing, policy engines, and guarded execution pipelines. The *semantic layer* applies inference-time guardrails, classifiers, and alignment mechanisms at the model invocation boundary. The role-structural layer is governed by the principle of *least agency*: constraining what an agent is permitted to decide, how that decision must be verified and by whom, prior to committing to it. This is distinct from the principle of least privilege that governs the resource-access layer. A system may satisfy least privilege while leaving least agency unbounded.

A production agentic system requires coherent work at all four layers. The architecture we propose in this paper develops an architectural pattern for the role-structural layer, advancing agentic AI system design practice and identifying the role-structural layer as a clear focus area for standardization work. Figure 1 provides an illustration of the logical view of the architecture.

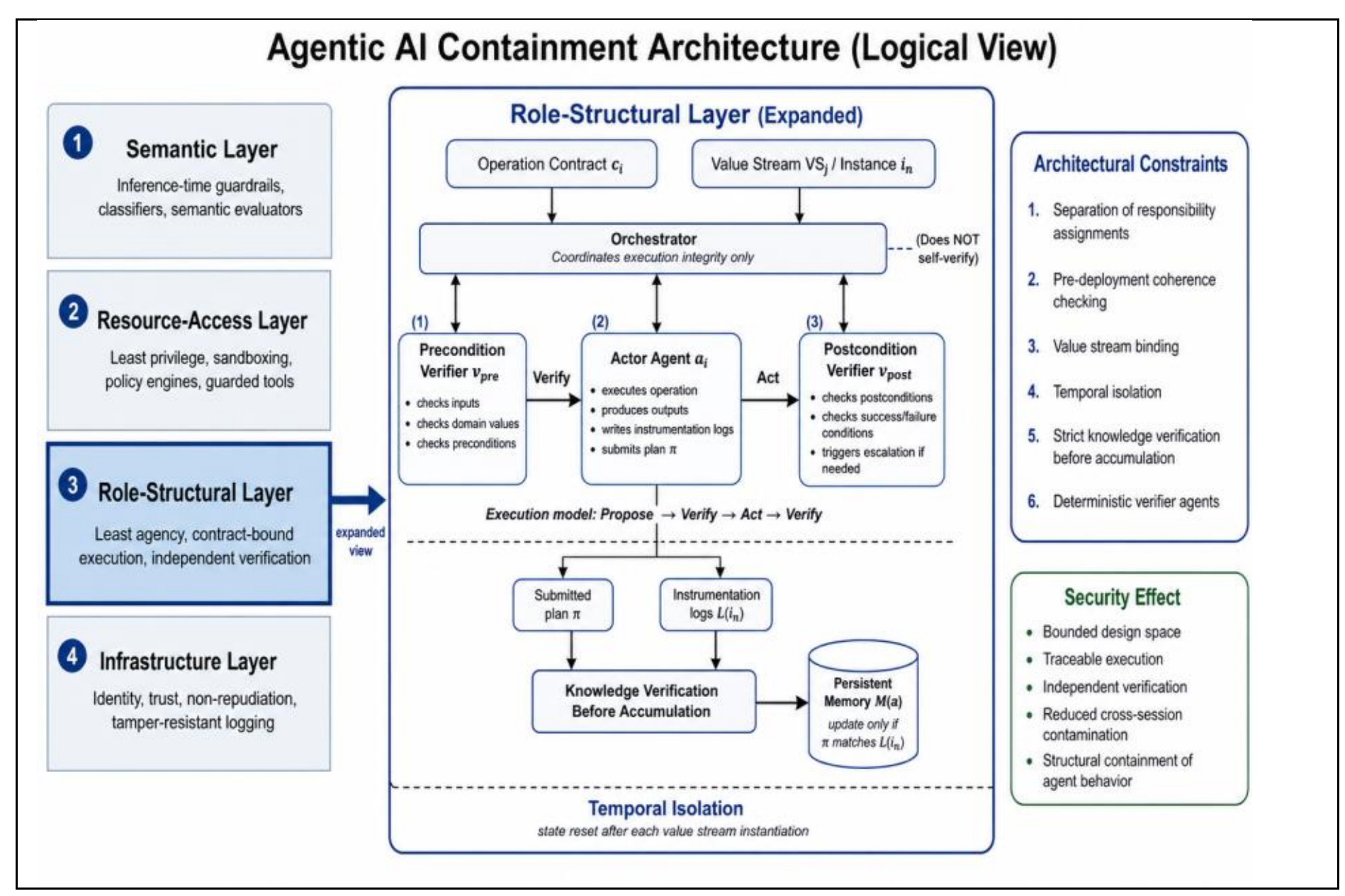

Figure 1. Agentic AI Containment Architecture

The architecture includes three main logical agentic roles. The precondition verifier role is responsible for verifying input formats, domain values, and preconditions, before an actor agent can be handed over execution control. The actor agent is responsible for carrying out the operation they are assigned, produce instrumentation logs, and produce a plan for evaluating the process the agent followed. The postcondition verifier role is responsible for verifying postconditions, success end conditions, failure end conditions, and triggering escalation if needed per bound contract. The orchestrator role is responsible for execution integrity across all roles. The architectural constraints outlined below impose a hard boundary on the set of possible designs. In return, compliant designs inherit desirable security hardening properties that derive directly from the interplay between the architectural constraints. We discuss these propositions in section 5.

In any domain of application, an operation is defined by its semantics. When the operation is carried out by a human (e.g. HR employee reviewing resumes), a combination of tacit knowledge and Standard Operating Procedures or business process normative specification is followed. This knowledge is the source of contractual obligations defining normative operation. The dominant pattern for ensuring alignment with the normative definition of an operation is Entry-Task-Validate-Exit. This pattern maps to verifying entry conditions before attempting an operation, executing the operation, verifying that the operation's outputs satisfy the expected postconditions, and exiting the flow after any remaining housekeeping. This paper ports this pattern into Agentic AI by treating AI model-based agents as proposers of decisions that must be verified before committing to, resulting in the pattern of Propose-Verify-Act-Verify.

The set of architectural constraints below bounds the alignment-by-design space in the agentic AI context. The result is contractual alignment as a verifiable operational outcome.

The *separation of responsibility assignments* constraint imposes a strict role separation between actor agents responsible for carrying out an operation in the business process and verifier agents responsible for checking preconditions, postconditions, submitted plans, etc. against the contracts and instrumentation logs. Verifier agents are bound to the contracts they verify. One verifier agent per contract. This ensures that only a specific contract with a globally unique hash can be admitted to the agent responsible for verifying it. A contract-agent mismatch immediately escalates to a human as it may signal a malicious intent. The contract to be verified by an agent must be fed to the agent as an input. The contract cannot be encapsulated by the agent verifying it. The deployment of the agent bound to a specific contract hash is traceable to a human or legal entity to guarantee non-repudiation.

The *pre-deployment coherence checking* constraint requires contract coherence checking across contracts and human approval before deployment in production. Coherence checking at runtime is also required for state-dependent checks that may induce contract coherence violations.

The architecture incorporates a *value stream binding* constraint mandating that each operation contract must be bound to each value stream that the operation is invoked/executed on. Each value stream must be uniquely identifiable. Each value stream execution instantiation must be uniquely identifiable and time-stamped. Whenever an operation is executed, the execution must be tagged with the value stream ID. When a value stream that invokes agentic operations is instantiated again, the agent is bound to the value stream instantiation it is executing in service of. If an operation behaves differently across value streams, then such operation must be modeled as separate operations with a distinct operation contract with appropriate binding to applicable value streams. This constraint strengthens the traceability and contextual integrity of agentic execution by ensuring that no operation exists in a vacuum; every action is anchored to a specific, uniquely identified business process (value stream) and a specific moment in time (value stream instance).

The *temporal isolation* constraint limits the lifetime of a value stream execution instantiation to a single traversal of the value stream (e.g. one complete resume review). Temporal isolation requires the implementation of a state reset mechanism to reset each agent's state on the value stream to the last known coherent state of the agent's persistent memory. When the value stream instantiation concludes successfully, or aborted, a new instantiation must be uniquely identified and started from scratch completely unaffected by prior runs. AI agents carrying out operations on the value stream lose the working memory of this particular instantiation's execution when the value stream is completed or aborted.

For continuous learning environments, the *strict knowledge verification before accumulation order* constraint allows the selective filtering needed for continuous learning autonomous systems. The strict order is enforced through instrumentation of individual actor agents. Actor agents, upon success, failure, or escalation, can construct the plan they followed and send the plan to the verifier agent. The verifier agent then retrieves the instrumentation logs to compare against the submitted plan by the actor agent. If the plan (not the actual data of the operation) matches the instrumentation logs, then the plan gets added to the agent's persistent memory. Otherwise, the plan is discarded per the temporal isolation constraint. This coordinated interplay between temporal isolation and selective knowledge augmentation protects the agent from drift as long-running execution and other multi-agent interactions feed into its execution context. This systemic risk is managed by the strict adherence to orderly knowledge augmentation over time when modifying the agent's persistent memory.

The *deterministic verifier for structural and process integrity properties* constraint mandates that only deterministic non-LLM verifier implementations be used for ensuring value stream binding, contract hash matching, schema conformance, state reset verification, temporal isolation, escalation trigger conditions, cross-layer coherence are indeed all and completely adhering to normative contracts and architectural constraints. This constraint enables replayability, auditability, and resistance to adversarial manipulation.

However, semantic content evaluation properties such as harm detection, contextual appropriateness, adversarial pattern recognition cannot be fully captured by deterministic specification. We propose augmenting deterministic verifiers with an upstream semantic verification layer, responsible for interpreting unstructured inputs, extracting relevant features, performing classification, and generating candidate judgements. This layer operates probabilistically and is optimized for flexibility and expressive understanding rather than strict correctness guarantees. In addition, an optional third layer may be introduced to provide semantic guardrails. This layer, which can include independent LLM-based evaluators or specialized classifiers, does not possess decision authority but instead generates risk signals, flags anomalies, and triggers escalations when semantic concerns are detected.

Together, these layers preserve the integrity of the containment architecture by isolating probabilistic reasoning from deterministic enforcement while still allowing the system to benefit from semantic capabilities where necessary. While it is theoretically possible for semantic verifiers to be exploited to pollute or hijack contract verifica-tion, the attack surface is meaningfully narrow due to the structured architectural constraints.

The six constraints above assume that each agent is uniquely identifiable within the execution context. Agentic identity design and implementation must guarantee attribition whether the deployment is internal, cross-organizational, or global.

Swapping a contract for a newer approved version requires a rebuild and redeployment of its verifier agent. The new hash of the contract will not be usable until built into the verifier, immediately triggering aborting execution until the mismatch is corrected. However, swapping a contract-agent pair has no ripple side effects or updates on other actor or verifier agents.

Sound execution of domain operations rests upon coherent contracts that define the alignment boundaries for execution. It is important to note that the system is only as reliable as its knowledge representation. Incomplete, outdated, or incorrectly modeled rules can propagate errors across the contract-driven AI system. This risk is already present in non-AI systems but magnified here. The combination of a grounded domain ontology and structured operation contracts enables the use of LLMs as a semantic-to-symbolic compilation layer for drafting machine-verifiable specifications. However, correctness is not derived from the LLM output itself, but from subsequent deterministic validation and human approval, preserving the integrity of the contract-verification architecture.

## 4 Operation Contract Discovery Process

Contracts are derived from the semantics of domain operations and must be expressed in a form that is both semantically grounded and structurally verifiable. This requirement is less burdensome than it may appear. Domain operation modeling [3] is already a prerequisite of any well-engineered system operating within business processes. In enterprise software development, identifying operation boundaries, input-output specifications, preconditions, postconditions, success and failure conditions is standard (not optional) systems analysis work that must be performed regardless of implementation technology. This work is already performed, to varying degrees of formality. The contribution of this section is to show that this pre-existing analytical work, when organized under the architectural constraints introduced in Section 3.2, directly yields machine-verifiable contracts without requiring a separate or novel modeling effort. The process below formalizes this mapping. Equally important, this process brings to the forefront whether an AI solution is recommended or needed to begin with to justify the cost and potential expansion of risk exposure.

1. Identify a specific use case. Select a business process suitable for AI augmentation, ensuring that the process has clear boundaries, stakeholders, and operational significance.
2. Articulate expected benefits. Define measurable outcomes aligned with strategic intent over a two- to four-quarter horizon, specifying how success will be evaluated in terms of performance, efficiency, risk reduction, or other relevant business metrics.

3. Assess data maturity and observability. Evaluate whether all required operation inputs and state variables are available, reliably captured, and observable at execution time. Where gaps exist, LLMs grounded in retrieval-augmented generation may be used to extract process activities into a structured Entry–Task–Validate–Exit representation for human validation, with particular attention to ensuring that all required state transitions can be observed and verified.
4. Express the use case semi-formally. Represent the process as a fully dressed, sea-level use case, including the main success path, alternate flows, failure paths, and escalation conditions, so that all relevant execution scenarios are explicitly captured.
5. Identify intelligent behaviors. Define the specific decision-making or interpretive responsibilities that AI agents will carry out within the use case, distinguishing these from purely mechanical or deterministic steps.
6. Decompose into operations. Break each identified intelligent behavior into atomic operations that have well-defined distinct responsibilities, boundaries, that can be independently verified, and operate over observable system state, ensuring that each operation corresponds to a unit of execution that can be governed by a contract.
7. Define operation contracts. For each operation, specify required inputs, optional inputs, preconditions, postconditions, invariants, failure end conditions, and special conditions such as termination, escalation, rollback, and human intervention, ensuring that all conditions are expressed in terms that are observable and computable from the system state.
8. Separate structural and semantic predicates. Classify each condition within the operation contract as either structural or semantic, where structural predicates are enforced by deterministic verifier agents and semantic predicates are delegated to a separately governed inference layer.
9. Validate contract completeness and coherence. Ensure that all execution paths defined in the use case are covered by operation contracts, that no conflicting or overlapping contracts exist, and that all predicates are grounded in the domain model and correspond to observable system state.

# 5 Threat Coverage Analysis

## 5.1 Notation and Shared Definitions

Let $\mathcal{D}$ denote the set of all designs compliant with the architecture. A design $d \in \mathcal{D}$ satisfies all six constraints simultaneously. Table 1 presents the constraint nomenclature used throughout the paper.

**Table 1.** Constraint Symbols.

| Symbol | Constraint |
|---|---|
| **S** | Separation of Responsibility Assignment |
| P | Pre-deployment Contract Coherence Checking |
| **V** | Value Stream Binding |
| **T** | Temporal Isolation |
| **K** | Strict Knowledge Verification Before Accumulation |
| D | Deterministic Verifier Agents |

$h(\cdot)$: a cryptographic hash function assumed collision-resistant
$c_i$: the contract governing operation $i$, with globally unique hash $h(c_i)$
$v_i$: the verifier agent bound to $c_i$
$a_i$: the actor agent executing operation $i$
$VS_j$: value stream $j$
$i_n$: the $n$-th instantiation of a value stream
$L(i_n)$: the instrumentation log produced during instantiation $i_n$
$\pi$: a plan submitted by an actor agent after execution
$M(a)$: the persistent memory of agent $a$

### 5.2 Hijacking via Contract Substitution

Defended by: Separation of responsibility assignments (**S**) + Deterministic verifier agents (**D**)

**Proposition 1.** *For any design $d \in \mathcal{D}$ satisfying S and D, no adversarially substituted contract $c' \neq c_i$ can be silently admitted to verifier agent $v_i$ without triggering escalation.*

**Reasoning.** By **S**, verifier agent $v_i$ is bound to exactly one contract $c_i$ with globally unique hash $h(c_i)$. The contract is supplied to $v_i$ as an external runtime input and cannot be encapsulated within $v_i$'s internal state. This externalization means $v_i$ does not own the contract it verifies, so the contract cannot be silently overwritten through state manipulation of $v_i$ itself.

By **D**, contract hash matching is implemented by a deterministic, non-LLM verifier. The verifier's behavior on any input is a fixed computable function and therefore cannot be redirected through adversarial prompting, jailbreaking, or semantic manipulation. **D** is necessary here: an LLM-based verifier nominally bound to $c_i$ could be induced through adversarial inputs to accept a substituted contract without triggering escalation. The non-LLM deterministic requirement eliminates this vector.

Suppose an adversary supplies contract $c'$ to $v_i$ where $c' \neq c_i$. By collision resistance of $h(\cdot)$, $h(c') \neq h(c_i)$. The deterministic verifier computes $h(c')$ and compares it against the registered $h(c_i)$. The inequality is detected. By **S**, a contract-agent hash mismatch is a designated escalation trigger, flagged as a signal of possible malicious intent.

Therefore, no adversarially substituted contract can transit $v_i$ without producing an escalation event. Silent admission is impossible under the conjunction of **S** and **D**.

### 5.3 Orchestrator Manipulation and Rogue Orchestration

Defended by: Separation of responsibility assignments (**S**) + Deterministic verifier agents (**D**)

**Architectural assumption.** The reasoning below assumes that instrumentation logs $L(i_n)$ are produced and maintained by a tamper-resistant mechanism outside the control of any individual agent, including the orchestrator. This is a necessary precondition for Case 3's detection argument. If the logging mechanism is itself compromised, the structural gap detection described in Case 3 cannot be relied upon. Formalizing tamper-proof logging as a verifiable architectural property is identified as a direction for future work.

**Proposition 2.** *For any design $d \in \mathcal{D}$ satisfying **S** and **D**, a compromised orchestrator agent cannot self-verify its own actions or suppress verification outcomes.*

**Reasoning.** By **S**, role assignments between actor agents and verifier agents are strictly separated. An orchestrator occupies an actor role. No agent may simultaneously hold actor and verifier roles; the two are mutually exclusive by architectural assignment. We consider all suppression strategies available to a compromised orchestrator $a_{orch}$:

*Case 1:* Direct self-verification. By **S**, $a_{orch}$ is not the verifier of any contract governing its own actions. The verifier $v_i$ for any operation $a_{orch}$ executes is a distinct, independently deployed agent. Self-verification is structurally unavailable under **S**.

*Case 2:* Verifier replacement. For $a_{orch}$ to substitute $v_i$ with a compromised verifier $v'_i$, it must deploy an agent bound to $h(c_i)$. Two sub-cases arise:

If $v'_i$ is bound to a different contract $c' \neq c_i$, the hash mismatch is detected by the deterministic verifier per **D** and triggers escalation per **S**. **D** is necessary here: an LLM-based replacement verifier nominally bound to $c_i$ could be induced to accept a mismatched contract through adversarial reasoning. The non-LLM deterministic requirement under **D** eliminates this vector. If $v'_i$ is bound to the identical contract $c_i$ and implements hash matching deterministically per **D**, then by definition it performs identical verification and not a rogue verifier. In both sub-cases $a_{orch}$ cannot deploy a replacement verifier that silently accepts a mismatched contract under the conjunction of **S** and **D**.

*Case 3:* Bypassing verification entirely. The Propose–Verify–Act–Verify pattern mandated by the architecture requires verification at each execution transition. Skipping a verification step means the corresponding postcondition check produces no entry in $L(i_n)$. Under the tamper-resistant logging assumption stated above, this structural gap in $L(i_n)$ is detectable by downstream verifiers and coherence checks, preventing silent bypass.

If the logging mechanism cannot be assumed tamper-resistant, Case 3's detection argument does not hold independently. In such deployments, logging integrity must be enforced through implementation-level controls outside the scope of these six constraints.

**Corollary 2.1.** By **S**, the blast radius of any individual verifier compromise is bounded to the single contract it governs. No verifier is shared across multiple contracts, so a compromised $v_i$ cannot propagate verification failures to operations governed by other contracts.

### 5.4 Cross-session State Poisoning and Memory Contamination

Defended by: Temporal isolation (**T**) + strict knowledge verification before accumulation (**K**)

**Architectural assumptions.** The reasoning below relies on two assumptions. First, that the state reset mechanism required by **T** is itself correctly implemented and verifiable. Second, that instrumentation logs $L(i_n)$ are tamper-resistant per the assumption stated in Proposition 2.

**Proposition 3.** *For any design $d \in \mathcal{D}$ satisfying **T** and **K**, no content from the execution of value stream instantiation $i_n$ can be written into the persistent memory $M(a)$ of any agent a without passing verifier-mediated knowledge screening, and no state from $i_n$ persists into $i_{n+1}$ through working memory, under the tamper-resistant logging and correct state reset assumptions above.*

**Reasoning.** We address the two contamination vectors separately.

*Vector 1: Working memory persistence across instantiations.*

By **T**, each value stream instantiation $i_n$ has a lifetime bounded to a single traversal. Upon conclusion or abort of $i_n$, the state reset mechanism restores each agent's state to its last known coherent state. Working memory accumulated during $i_n$ is discarded. By construction, $i_{n+1}$ begins from this coherent baseline, unaffected by $i_n$'s execution context. Under the correct state reset assumption, working memory contamination across instantiations is structurally prevented by **T**.

*Vector 2: Persistent memory contamination via knowledge accumulation.*

By **K**, for any plan $\pi$ constructed by actor agent a during $i_n$, the following strict ordering is enforced before $\pi$ is eligible for persistence in $M(a)$:

1. a submits $\pi$ to verifier $v_i$
2. $v_i$ retrieves $L(i_n)$, the instrumentation log of $i_n$

3. $v_i$ checks that π matches $L(i_n)$
4. Only upon successful match is π admitted to $M(a)$

For an adversary to inject malicious plan $\pi^*$ into $M(a)$, $\pi^*$, it must match $L(i_n)$. Matching $L(i_n)$ requires that the actual execution during $i_n$ followed $\pi^*$, meaning the adversarial behavior was already executed in a manner consistent with the instrumentation log. This detection argument is not fully independent: it relies on the broader PVAV execution model functioning correctly across the system, including the defenses established in Propositions 1, 2, and 5. Therefore, Vector 2's defense is a component of the system-level containment architecture rather than a self-contained guarantee.

**Conjunction.** **T** eliminates cross-session working memory contamination. K eliminates unverified persistent memory accumulation within a session. Together, under the stated assumptions, they close both structural contamination vectors, establishing that $M(a)$ can only grow through verifier-confirmed, log-consistent knowledge increments.

### 5.5 Unsanctioned Goal Generalization

Defended by: Value stream binding (**V**) + pre-deployment coherence checking (**P**) + Deterministic Verifier Agents (**D**)

By anchoring every operation to a specific, uniquely identified value stream instantiation, the architecture prevents agents from "generalizing" their behavior across contexts in ways that weren't sanctioned. An operation that behaves differently across value streams must be modeled as a separate operation with its own contract. The result is forcing that behavioral divergence into the open at design time rather than letting it emerge silently at runtime.

**Proposition 4.** *For any design $d \in \mathcal{D}$ satisfying **V**, **P**, and **D**, an agent cannot generalize sanctioned behavior from value stream $VS_j$ to value stream $VS_k$ ($j \neq k$) without either explicit contract authorization or detection at runtime or design time*.

**Reasoning.** By **V**, every operation o is bound to the specific value stream(s) it is sanctioned to serve. Each operation execution is tagged with a unique value stream ID and instantiation timestamp. If operation o exhibits meaningfully different behavior across $VS_j$ and $VS_k$, it must be modeled as two distinct operations $o_j$ and $o_k$ with separate contracts $c_j$ and $c_k$, each bound to its respective value stream.

Suppose agent a generalizes behavior from $VS_j$ to $VS_k$ without distinct authorization. Two sub-cases arise:

*Sub-case 1:* At runtime. The generalized execution of $o$ under $VS_k$ carries a $VS_k$ execution tag. The verifier $v_j$, bound to $c_j$, checks that the operation's value stream binding matches $c_j$'s specification. A $VS_k$ tag under $c_j$ is a value stream binding violation. By **D**, this check is implemented by a deterministic, non-LLM verifier and therefore cannot be bypassed through adversarial prompting or semantic manipulation. The violation is detected and escalation is triggered under **V** and **D**.

*Sub-case 2:* At design time. If the design encodes *o* as a single operation spanning both *VS_j* and *VS_k* without distinct contracts, this cross-stream behavioral conflation constitutes a contract coherence violation under **V**. By **P**, contract coherence across all contracts is checked before deployment and requires human approval. This violation is surfaced and must be resolved before the design can be deployed.

In both sub-cases, unsanctioned cross-value-stream generalization is either escalated at runtime (**V** + **D**) or blocked at deployment (**P**).

### 5.6 Emergent Agent Collusion

Defended by: Separation of responsibility assignments (**S**) + value stream binding (**V**) + Strict knowledge verification before accumulation (**K**)

**Architectural assumption.** The detection argument in Case 2 relies on instrumentation logs *L(i_n)* being tamper-resistant per the assumption stated in Proposition 2.

**Proposition 5.** *For any design d ∈ 𝒟 satisfying S, V, and K, agents cannot coordinate outside their sanctioned interaction paths without producing detectable anomalies in instrumentation logs L(i_n), under the tamper-resistant logging assumption above.*

**Reasoning.** By **V**, every legitimate operation execution during instantiation *i_n* is tagged with the value stream ID of *i_n* and recorded in *L(i_n)*. All sanctioned inter-agent interactions are operations defined in contracts and therefore appear as tagged entries in *L(i_n)*. We consider all coordination strategies available to colluding agents:

**Case 1: Coordination through a sanctioned operation.** If the interaction $\iota$ between agents *a_1* and *a_2* is a contract-defined operation, it is tagged in *L(i_n)* and subject to verification. This interaction is architecturally visible and does not constitute covert collusion.

**Case 2: Coordination through an unsanctioned operation.** If $\iota$ is not defined in any contract, it carries no value stream tag and produces no corresponding entry in *L(i_n)*. By **K**, when verifier *v_i* reconciles the plan $\pi$ submitted by an actor agent against *L(i_n)*, the absence of a log entry for $\iota$ constitutes a plan-log mismatch, triggering plan rejection. Alternatively, if $\iota$ is embedded within a submitted plan $\pi$, the verifier checking $\pi$ against *L(i_n)* will find an operation in $\pi$ with no corresponding log entry, again producing a mismatch.

By **S**, no actor agent can suppress its own verification, ensuring that the reconciliation step under **K** cannot be bypassed by a colluding actor.

**Corollary 5.1.** By **S**, the blast radius of any individual verifier compromise is bounded to the single contract it governs. No verifier is shared across multiple contracts, and no actor agent controls its own verifier. A compromised verifier $v_i$ can therefore affect only the operations governed by $c_i$ and cannot propagate verification failures to operations governed by other contracts.

### 5.7 Specification Gaming and Loophole Exploitation

Defended by: Pre-deployment coherence checking (**P**)

Contract coherence is verified before deployment, not discovered at runtime. This is a direct defense against agents finding and exploiting underspecified or contradictory rules. However, there is a residual risk because the system is only as reliable as its knowledge representation, so incomplete modeling remains an attack surface.

**Proposition 6.** *For any design $d \in \mathcal{D}$ satisfying P, no loophole detectable by static cross-contract coherence analysis can survive to production deployment.*

**Reasoning.** Define a loophole $\ell$ as a behavioral path that: (i) satisfies the letter of at least one individual contract $c_i$ in isolation, but (ii) violates the normative intent of the contract set $C = \{c_1, ..., c_n\}$ through exploitation of a contradiction between $c_i$ and $c_j$ ($i \neq j$), or a coverage gap that admits unsanctioned behavior not prohibited by any single contract.

By **P**, the contract set $C$ undergoes coherence checking as a whole before deployment. Human approval is additionally required. Let *Detectable*($C$) denote the set of loopholes detectable by static cross-contract coherence analysis on $C$. For any $\ell \in$ *Detectable*($C$), the coherence check surfaces $\ell$ before deployment. By **P**, deployment is blocked until all detected violations are resolved. Therefore, no $\ell \in$ *Detectable*($C$) survives to production for any $d \in \mathcal{D}$.

### 5.8 Semantic Adversarial Content Attacks

Defended by: Deterministic Verifier Agents (**D**)

**Proposition 7.** *For any design $d \in \mathcal{D}$ satisfying D, the semantic attack surface is structurally contained to the inference layer IL(d) at the model invocation boundary. The architecture provides no guarantee on the effectiveness of IL(d); whether residual semantic misalignment risk falls within a deployment-specific acceptable bound t is exclusively a function of IL(d)'s design and operational strength.*

**Reasoning.** *Part 1: Structural containment of the semantic attack surface to IL(d).* By **D**, semantic content evaluation such as harm detection, adversarial pattern recognition, and contextual appropriateness assessment is explicitly excluded from the deterministic verifier boundary. Deterministic verifiers are non-LLM implementations that process only structured artifacts: contract hashes, schema conformance results, value stream binding tags, and state reset verification outcomes. They cannot be manipulated through adversarial natural language inputs because they do not process natural language.

All semantic content transits through *IL(d)* at the model invocation layer before it can influence agent behavior. There is no architectural path by which semantic content reaches the contract verification path directly. For any adversarial semantic input *x*, the only channel through which *x* can induce misalignment in the structural verification layer is through *IL(d)* first. The deterministic verifier path is semantically opaque and structurally isolated from *x* by **D**. **D** therefore guarantees that the semantic attack surface is contained to *IL(d)* and cannot exceed it.

*Part 2: Scope of the architecture's semantic safety claim.* Let $t \in [0, 1]$ be the deployment-specific bound of acceptable residual misalignment risk from semantic attacks, established by the deploying organization. Let *R_sem*(*d*) denote the residual misalignment risk from semantic attacks in design *d*, defined as the probability that *IL(d)* fails to correctly intercept a semantic attack drawn from the deployment's operational threat environment.

The architecture's semantic safety boundary is that the attack surface that *R_sem*(*d*) measures cannot exceed *IL*(*d*). Whether *R_sem*(*d*) ≤ *t* is achieved is entirely an *IL*(*d*) design and operational problem.

*Part 3: Necessity of IL(d) strength.* If *R_sem*(*d*) > *t*, the probability of semantic misalignment exceeds the deployment's acceptable bound regardless of the strength of constraints **S**, **P**, **V**, **T**, and **K**. Those constraints govern structural and process integrity and provide no coverage of semantic content evaluation. No combination of structural constraints can substitute for a correctly specified and operationally maintained *IL*(*d*).

**Residual risk: semantic-to-structural boundary contamination.** Part 1 establishes that the deterministic verification path cannot be directly manipulated through semantic content. However, the boundary between semantic inputs and structured outputs is not fully impermeable in practice. Adversarial semantic content processed by an actor agent could influence the structured artifacts that agent produces. For example, causing a parsing agent to emit a schema-valid but adversarially constructed output that exploits edge cases in deterministic schema validation. In such cases, the attack path runs from semantic input through *IL*(*d*), into actor-produced structured output, and then into the deterministic verification path in a form that is structurally non-compliant but potentially boundary-exploiting.

**Corollary 7.1.** The attack surface exposed to semantic attacks is minimal under **D** and cannot be further reduced within the architecture unless deterministic verification is a blanket constraint for all verifiers. Reducing *R_sem*(*d*) is therefore exclusively an *IL*(*d*) design and operational problem, not an architectural one.

## 6 Case Study: Resume Screening Value Stream

To ground the proposed architecture in a concrete operational setting, this section instantiates the constraints within a resume screening workflow. Resume screening is a suitable domain because it is both operationally well-defined and subject to regulatory, fairness, and auditability requirements, while also being exposed to adversarial inputs and model-induced bias. Let (*VSscreen*) denote the resume screening value stream, where each instantiation ($i_n$) corresponds to the evaluation of a single candidate against a specific job requisition. Each instantiation is uniquely identified by a value stream identifier, an instance identifier, and a timestamp. The flow progresses from consuming inputs including the candidate resume, job description, and applicable screening policies to a terminal state producing one of three outcomes: advance, reject, or escalate for human review.

The value stream is decomposed into a sequence of domain operations, including resume parsing, criteria extraction, candidate evaluation, policy compliance checking, and decision formation. Each operation ($o_i$) is governed by a contract ($c_i$) with a globally unique hash ($h(c_i)$), executed by an actor agent ($a_i$) and verified by a corresponding verifier agent ($v_i$). In accordance with the separation of responsibility assignments constraint, actor and verifier roles are strictly disjoint. Verifier agents receive their associated contracts as external inputs and are responsible for validating preconditions, postconditions, and execution traces against the contract specifications. For example, the candidate evaluation operation enforces that only permitted features such as skills, experience, and certifications are used in scoring, while explicitly prohibiting the use of protected attributes. The verifier checks both schema conformance and feature attribution, ensuring that outputs adhere to the normative constraints defined in the contract.

Execution within a value stream instantiation follows the Propose–Verify–Act–Verify pattern. An actor agent first proposes a plan of execution, which is validated by the verifier against contract preconditions. Upon successful validation, the actor executes the operation, producing outputs and instrumentation logs. The verifier then performs postcondition checks, including consistency between the submitted plan and the recorded execution trace. Any deviation from contract specifications or inconsistencies in execution result in escalation, preventing silent failure modes. This pattern enforces aligned execution not through probabilistic guarantees but through deterministic validation of structural properties.

Temporal isolation ensures that each value stream instantiation is executed independently, with working memory reset upon completion or abort. This prevents cross-candidate contamination, a known risk in adaptive systems where prior interactions may influence subsequent decisions. Persistent memory updates are governed by the strict knowledge verification before accumulation constraint, which requires that any plan proposed by an actor agent must be reconciled against instrumentation logs by a verifier before being admitted into persistent memory. As a result, only execution paths that are both contract-compliant and log-consistent contribute to long-term system behavior.

The architecture also provides structural defenses against common classes of adversarial behavior within this workflow. Prompt injection attempts embedded within resume content, such as instructions to override evaluation criteria, are constrained by the contract-bound verification process, which enforces feature usage and output schema independently of the model's semantic interpretation. Attempts to introduce bias through explicit use of protected attributes are constrained by contract prohibitions and deterministic verification of feature attribution. Implicit bias through proxy features remains a semantic concern delegated to the inference layer per Proposition 7. Orchestrator-level manipulation is mitigated by the separation of actor and verifier roles, ensuring that no component can self-validate its actions. Additionally, temporal isolation and controlled knowledge accumulation prevent the propagation of malicious or unintended behavior across value stream instantiations.

Each instantiation produces a comprehensive instrumentation log, including contract references, execution traces, and verification outcomes. This enables full auditability of screening decisions, supporting regulatory compliance and post-hoc analysis. Importantly, the architecture makes explicit its boundaries: semantic risks, such as subtle bias not captured by structural constraints, remain dependent on the effectiveness of the inference layer, while the correctness of the system is contingent on the completeness and accuracy of the contract specifications. These limitations are not eliminated but are explicitly bounded and exposed.

This case study demonstrates that the proposed architecture enforces structural integrity as an emergent property of constrained execution, while bounding semantic risk to the inference layer.

## 7 Conclusion

The Agent Containment Architecture proposed in this work operationalizes this position through a set of six interacting constraints that collectively enforce a Propose–Verify–Act–Verify execution model, ensuring that every operation is both contractually defined and independently verified. The result is a system in which structural integrity is a verifiable outcome of constrained design and observable execution, with semantic risk explicitly bounded and delegated to the inference layer. The threat coverage analysis and propositions provide reasoned arguments for the constraints by mapping attack classes to activated constraints. The constraint view is complemented by an explicit separation between structural integrity and semantic safety with appropriate mitigations and limits for each.

Future research directions include formalizing coherence checking methodologies, specifying verifiable governance models for contract lifecycle management, and empirical validation through prototype systems and adversarial testing scenarios is also necessary to quantify the practical effectiveness and performance trade-offs of the proposed constraints in production environments.

## 8 Declaration on Generative AI

Generative AI was used in the preparation of this work for researching related work, tightening prose to observe the page limit, proofreading, and checking for gaps in reasoning arguments.